\documentclass[aps,prd,preprint,tightenlines,twocolumn,nofootinbib,showpacs,fixfloat]{revtex4-1}
\usepackage{epsfig}
\usepackage{graphicx}% Include figure files
\usepackage{dcolumn}% Align table columns on decimal point
\usepackage{bm}% bold math
\usepackage{overpic}
\usepackage{subfigure}
\usepackage{float}
\usepackage{natbib}
\begin{document}
%\preprint{BESIII note NO.}

\title{\boldmath Precision Measurements of ${\cal B}[\psi(3686) \to \pi^+\pi^- J/\psi]$ and 
	${\cal B}[J/\psi\to l^+l^-]$}

\author{
\begin{center}
\begin{small}
M.~Ablikim$^{1}$, M.~N.~Achasov$^{7,a}$, O.~Albayrak$^{4}$, D.~J.~Ambrose$^{40}$, F.~F.~An$^{1}$, Q.~An$^{41}$, J.~Z.~Bai$^{1}$, R.~Baldini Ferroli$^{18A}$, Y.~Ban$^{27}$, J.~Becker$^{3}$
% ...
	\\
	\vspace{0.2cm}
(BESIII Collaboration)\\
	\vspace{0.2cm} {\it
		$^{1}$ Institute of High Energy Physics, Beijing 100049, People's Republic of China\\
% ...
$^{3}$ Bochum Ruhr-University, D-44780 Bochum, Germany\\
			$^{4}$ Carnegie Mellon University, Pittsburgh, Pennsylvania 15213, USA\\
% ...
$^{7}$ G.I. Budker Institute of Nuclear Physics SB RAS (BINP), Novosibirsk 630090, Russia\\
% ...
$^{18}$ (A)INFN Laboratori Nazionali di Frascati, I-00044, Frascati, Italy; (B)INFN and University of Perugia, I-06100, Perugia, Italy\\
% ...
$^{27}$ Peking University, Beijing 100871, People's Republic of China\\
% ...
$^{40}$ University of Rochester, Rochester, New York 14627, USA\\
			$^{41}$ University of Science and Technology of China, Hefei 230026, People's Republic of China\\
% ...
$^{a}$ Also at the Novosibirsk State University, Novosibirsk, 630090, Russia\\
% ...
	}
\vspace{0.4cm}
\end{small}
\end{center}
}

\date{\today}

\begin{abstract}
Based on $(106.41 \pm 0.86)\times 10^{6}$ $\psi(3686)$ events
collected with the BESIII detector at the BEPCII collider, the
branching fractions of $\psi(3686) \to \pi^+\pi^- J/\psi$ , $J/\psi
\to e^+e^- $, and $J/\psi \to \mu^+\mu^-$ are measured.  We obtain
${\cal B}[\psi(3686) \to \pi^+\pi^-J/\psi]=(34.98\pm 0.02\pm 0.45)\%$,
${\cal B}[J/\psi \to e^+e^-] = (5.983 \pm 0.007 \pm 0.037 )\%$ and
${\cal B}[J/\psi \to \mu^+\mu^-] = (5.973 \pm 0.007 \pm 0.038)\%$.
The measurement of ${\cal B}[\psi(3686) \to \pi^{+}\pi^{-}J/\psi]$
confirms the CLEO-c measurement, and is apparently larger than the
others.  The measured $J/\psi$ leptonic decay branching fractions
agree with previous experiments within one standard deviation. These
results lead to ${\cal B}[J/\psi \to l^+l^-] = (5.978 \pm 0.005 \pm
0.040)\%$ by averaging over the $e^{+}e^{-}$ and $\mu^{+}\mu^{-}$
channels and a ratio of ${\cal B}[J/\psi \to e^+e^-] / {\cal B}[J/\psi
\to \mu^+\mu^-] = 1.0017 \pm 0.0017 \pm 0.0033$, which tests $e$-$\mu$
universality at the four tenths of a percent level.  All the
measurements presented in this paper are the most precise in the world
to date.
\end  {abstract}

\maketitle
\section{\bf INTRODUCTION}
%\bigskip

Since the discovery almost four decades ago of the first charmonium
state, the $J/\psi$~\cite{ref:jpsi}, the states that have been studied
the most among the various conventional charmonium states found have
been the $J/\psi$ and $\psi(3686)$.  However, the largest branching
fraction in $\psi(3686)$ decays, ${\cal
B}[\psi(3686)\to\pi^{+}\pi^{-}J/\psi] ({\cal B}_{\pi\pi\psi})$ still
remains interesting both experimentally and theoretically.  On the
experimental side, the mass recoiling against the dipion system
$(M^{\rm rec.}_{\pi^{+}\pi^{-}})$ of this common decay mode can be used to
identify $J/\psi$ decays.  This makes ${\cal B}_{\pi\pi\psi}$ crucial
for the relevant measurements in charmonium decays and searching for
new particles, such as invisible particles in $J/\psi$ decays, as well
as the measurements of charmonium production rates in higher energy
collisions.  Because of its large size, the branching fraction, ${\cal
B}_{\pi\pi\psi}$, also imposes a limit on the rest of the decay
channels of $\psi(3686)$.  On the theoretical side, the transition
$\psi(3686) \to \pi^+\pi^- J/\psi$ relates to the interaction between
heavy quarks and gluons as well as hadronization, providing an
excellent testing ground for some theoretical predictions such as the
QCD multipole expansion~\cite{ref:multipole} and chiral
symmetry~\cite{ref:cahn}.

${\cal B}_{\pi\pi \psi}$, however, has changed dramatically in the
last decades~\cite{MARK1-ppj, DASP-ppj, E760-ppj, bes2scan,
cleo-ppj}. For example, the most recent result from CLEO-c, ${\cal
B}_{\pi\pi\psi}$=$(35.04\pm 0.8)\%$~\cite{cleo-ppj}, is apparently
larger than the former most precise result $(32.3\pm 1.4)\%$ from
BESII~\cite{bes2scan}. The situation, thus, demands additional, high
precision measurements of ${\cal B}_{\pi\pi\psi}$. The data sample of
$\psi(3686)$ collected with the BESIII detector, which is the world's
largest such sample, makes it possible to remeasure ${\cal
B}_{\pi\pi\psi}$ and clarify the discrepancy.

Similar to the transition $\psi(3686) \to \pi^+\pi^- J/\psi$, $J/\psi
\to e^+e^-$ and $\mu^+\mu^-$ are often used to identify the $J/\psi$
experimentally for they are the two largest and cleanest decay modes
of $J/\psi$. The branching fractions for the leptonic decays $J /\psi
\to e^+e^-~({\cal B}_{ee})$ and $J/\psi \to \mu^+\mu^-~({\cal
B}_{\mu\mu})$ are fundamental parameters of the $J/\psi$ resonance,
and hence of general interest. The process of a vector charmonium
decaying into a lepton pair is thought to occur through the
annihilation of the $c \bar c$~pair into a virtual photon, and thereby
is related to the $c \bar c$~wave function overlap at the origin,
which plays a direct role in potential models~\cite{potentials}.
%Then ${\cal B}_{ll}$ is also be used in the normalization
%of charmonia decaying to light hadrons with the assumption that these
%decays are through the same hardcore mechanism, i.e. an annihilation
%of the $c \bar c$~pair into a virtual photon.
Furthermore, the ratio ${\cal B}_{ee}/{\cal B}_{\mu\mu}$ provides a
test of lepton universality. The standard model predicts exact
lepton university for $ee$ and $\mu\mu$, and any deviation from unity
will indicate possible new physics effects or new decay mechanisms for
$J/\psi$ to $l^+l^-$, where $l$ may be either $e$ or $\mu$.  Also, as
the branching fraction of $J/\psi\to l^+l^-$ (${\cal B}_{ll}$) is important in the determination of the $J/\psi$
leptonic and total widths, ($\Gamma_{ee}$ and $\Gamma_{\rm
tot}$)~\cite{babar}, its precision is important for their
uncertainties.

%The current status of ${\cal B}_{ll}$ measurement is that 
${\cal B}_{ee}$ and ${\cal B}_{\mu\mu}$ have been measured to be
approximately equal, as expected from lepton universality combined
with a negligible phase space correction.
%, at $\sim5.9\%$. 
A relative
precision of $1\%$ on both ${\cal B}_{ee}$ and ${\cal B}_{\mu\mu}$
has been achieved through an average~\cite{pdg2012} over measurements,
which are dominated by the results from CLEO-c~\cite{cleo:ppjll} and
BESI~\cite{bes:jll}.

This paper describes the measurement of the branching fraction ${\cal
B}_{\pi\pi\psi}$, as well as ${\cal B}_{ll}$ via the decay $\psi(3686)
\to \pi^+\pi^- J/\psi$.  Measuring ${\cal B}_{ll}$ via $\psi(3686) \to
\pi^+\pi^- J/\psi$ has the advantage that there is no interference
with Bhabha or dimuon production, that would need to be considered in
measurements via direct $J/\psi$ production and decay in an
electron-positron collider.

Our overall analysis procedure is as follows. The observed number of
events, $N_{\pi\pi J/\psi}$ and $N_{ll}$ ($ll$ represents
$\pi^{+}\pi^{-}l^{+}l^{-}$ final states), are extracted by fitting to
data distributions or counting the signal candidate events
directly. The corresponding acceptances, $\epsilon_{\pi\pi J/\psi}$
and $\epsilon_{ll}$, are calculated based on Monte Carlo (MC)
samples.  Then ${\cal B}_{\pi\pi \psi}$ is calculated with the
equation
\begin{eqnarray}
{\cal B}_{\pi\pi \psi}&=& 
\frac{N_{\pi\pi \psi}}{\epsilon_{\pi\pi \psi}\times N_{\rm tot}}~,
\label{equation-ppj}
\end{eqnarray}
where $N_{\rm tot}$ is the number of $\psi(3686)$ events. 
${\cal B}_{ll}$ is calculated with
\begin{eqnarray}
{\cal B}_{ll}
&=& \frac{{\cal B}_{\pi\pi\psi} \times {\cal B}_{ll}  } { { \cal B}_{\pi\pi\psi} }
=\frac{ N_{ll} / (\epsilon_{ll} \times N_{\rm tot})  }
{ N_{\pi\pi \psi} /( \epsilon_{\pi\pi \psi} \times N_{\rm tot})  }\nonumber\\
&=&
\frac{ N_{ll} / \epsilon_{ll}  }
{ N_{\pi\pi \psi} / \epsilon_{\pi\pi \psi}  }
~.
\label{equation-ppjll}
\end{eqnarray}
Here it should be noted that Eq.~\ref{equation-ppjll} is
independent of the number of $\psi(3686)$ events, which is one
of the major sources of systematic uncertainties in the determination
of ${\cal B}_{\pi\pi\psi}$.

\section{\bf BEPCII and BESIII}
%\bigskip

BESIII/BEPCII, described in detail in Ref.~\cite{ref:bes3}, is a major
upgrade of the BESII detector and the BEPC accelerator~\cite{ref:bes2}
for studies of hadron spectroscopy and $\tau$-charm
physics~\cite{ref:bes3physics}.  The design peak luminosity of the
double-ring $e^+e^-$ collider, BEPCII, is $10^{33}$ cm$^{-2}$s$^{-1}$
at a beam current of 0.93 A.

The BESIII detector with a geometrical acceptance of 93\% of 4$\pi$,
consists of the following main components: 1) a main drift chamber
(MDC) equipped with 6796 signal wires and 21884 field wires arranged
in a small cell configuration with 43 layers working in a gas mixture
of He (40\%) and $\rm{C}_3\rm{H}_8$ (60\%). The single wire resolution
on average is 135 $\mu$m, and the momentum resolution for charged
particles in a 1 T magnetic field is 0.5\% at 1 GeV/$c$; 2) an
electromagnetic calorimeter (EMC) made of 6240 CsI (Tl) crystals
arranged in a cylindrical shape plus two end-caps. The energy
resolution is 2.5\% in the barrel and 5\% in the end-caps at 1.0 GeV;
the position resolution is 6 mm in the barrel and 9 mm in the end-caps
at 1.0 GeV; 3) a Time-Of-Flight system (TOF) for particle
identification with a cylindrically shaped barrel portion, made with
two layers with 176 pieces of 5 cm thick, 2.4 m long plastic
scintillators in each layer, and end-caps each with 96 fan-shaped, 5
cm thick, plastic scintillators.  The time resolution is 80 ps in the
barrel, and 110 ps in the end-caps, corresponding to a K/$\pi$
separation at the $2\sigma$ level up to about 1.0 GeV/$c$; 4) a muon
chamber system (MUC) made of 1000 m$^2$ of Resistive Plate Chambers
(RPC) arranged in 9 layers in the barrel and 8 layers in the
end-caps. The position resolution is about 2 cm.

\section{\bf Event selection\label{evt-selection}}
%\bigskip

The data sample used for this analysis consists of $(106.41 \pm 0.86)
\times 10^{6}$ $\psi(3686)$ decays produced at the resonance
peak~\cite{ref:psiptotnumber} and an additional 44 pb$^{-1}$ of data
collected at $\sqrt{s} = 3.65$ GeV to determine the non-resonant
background contributions. A MC sample of $106\times 10^{6}$
$\psi(3686)$ inclusive decay events is used to obtain the detection
efficiencies as well as to estimate the backgrounds.  This sample is
generated with KKMC~\cite{ref:kkmc} and EvtGen~\cite{ref:bes3gen} for
decays with known branching fractions~\cite{pdg2006}, or by LundCharm
\cite{ref:lundcharm} for unmeasured decays.  The signal process of
$\psi(3686) \to \pi^{+}\pi^{-}J/\psi$ is generated according to the
formulas and measured results in Ref.~\cite{ref:bes1ppjdistribution},
which takes the small $D$-wave contribution into account.  The
$J/\psi \to l^{+}l^{-}$ processes are generated with an angular
distribution of $(1 + \cos^{2} \theta_{l})$, where $\theta_{l}$ is the
lepton angle relative to the beam line in the $J/\psi$ rest frame, and
PHOTOS~\cite{photos} is used for the final state radiation.  These MC
events are then processed with the detector simulation package based on
GEANT4~\cite{ref:geant4}.

In order to suppress tracks due to cosmic rays and beam associated
events, charged tracks are required to pass within $\pm10$ cm of the
run-by-run determined interaction point along the beam direction and
within $1$ cm of the beam line in the plane perpendicular to the beam.
To guarantee good agreement between data and MC simulation, all the
charged tracks must lie in the barrel region, i.e., $\vert
\cos\theta\vert<0.8$, where $\theta$ is the polar angle with respect
to the positron beam direction.

\begin{figure}[htbp]
\centering
\includegraphics[width=8cm,height=6.5cm]{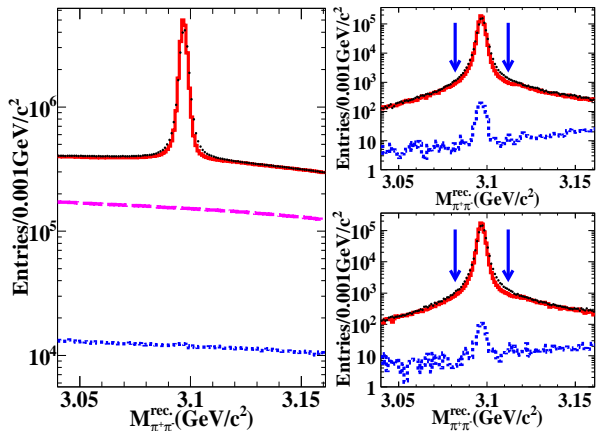}
\caption{(left) Distributions of $M_{\pi^{+}\pi^{-}}^{\rm rec.}$,
  where candidate events are represented by black dots, the
  non-$\pi^{+}\pi^{-}J/\psi$ decays of $\psi(3686)$ background by the
  purple long dashed line, the scaled continuum by the blue dashed
  dotted line, and the $\psi(3686)$ inclusive MC plus the scaled
  continuum and non-$\pi^{+}\pi^{-}J/\psi$ background by the red
  histogram.  Distributions of $M_{\pi^{+}\pi^{-}}^{\rm rec.}$ (top
  right) $J/\psi \to e^{+}e^{-}$ and (bottom right) $J/\psi \to
  \mu^{+}\mu^{-}$ candidate events, where only total backgrounds are
  shown with blue dash-dotted lines. The arrows indicate the mass
  windows to count the number of signal candidates.  }
\label{fig:ppjxx}
\end{figure}

To identify $\pi^{+}\pi^{-}J/\psi$ candidates,
$M_{\pi^{+}\pi^{-}}^{\rm rec.}$ is determined for all pairs of charged
tracks of opposite charge with momentum less than 450 MeV$/c$, that
are assumed to be pions, and all the combinations with
$M_{\pi^{+}\pi^{-}}^{\rm rec.}$ near the $J/\psi$ peak are kept
($[3.04,3.16]$ GeV/$c^{2}$).  The $(n)\gamma J/\psi$ backgrounds with
an electron-positron pair converted from a photon are removed by
requiring the cosine of the angle between the two charged tracks be
less than 0.95.  $N_{\pi\pi J/\psi}$ is determined from a fit to the
distribution of $M^{\rm rec.}_{\pi^{+}\pi^{-}}$.  The left plot in
Fig.~\ref{fig:ppjxx} shows the distribution of $M^{\rm
rec.}_{\pi^{+}\pi^{-}}$ for data, non-$\pi^{+}\pi^{-}J/\psi$ decays,
the scaled continuum events, and the sum of the signal from MC simulation and all backgrounds.  
Note that the mass resolutions
of data (black dots) and MC simulation (red histogram) are different,
which is considered in the following sections.

\begin{figure}[htbp]
\centering
\includegraphics[width=7.5cm,height=8.5cm]{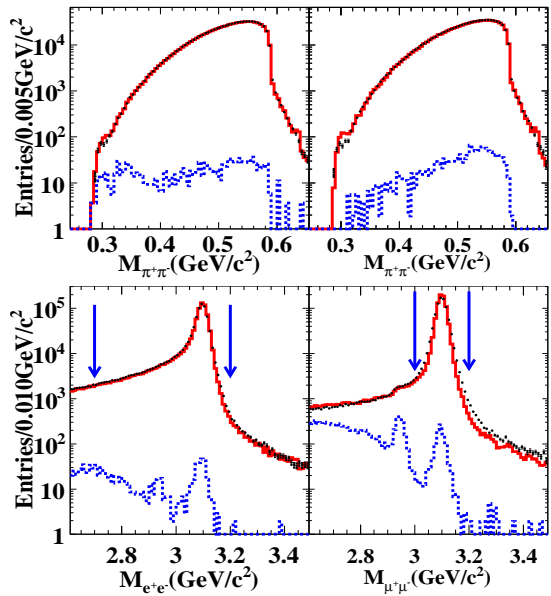}
\caption{ Distributions of $\psi(3686)\to\pi^{+}\pi^{-}J/\psi,~J/\psi
\to e^{+}e^{-}$ (left) and $J/\psi \to \mu^{+}\mu^{-}$ (right)
candidate events in the $\psi(3686)$ data (black dots with error
bars), MC simulation of signal plus background (red solid histogram),
and backgrounds (blue dashed dotted line).  The top panel shows
distributions of the dipion invariant mass, and the bottom panel shows
the dilepton invariant mass.  The arrows shown in each plot indicate
nominal selection criteria, which are applied for the other plots in
the figure.}
\label{fig:ppjll}
\end{figure}

For the selection of $\pi^{+}\pi^{-}l^{+}l^{-}$ candidates, the pion
pair is identified in the same way as for $\pi^{+}\pi^{-}J/\psi$.
When multiple entries occur, the one with the minimum $\vert
M^{\rm rec.}_{\pi^{+}\pi^{-}} -m_{J/\psi}\vert$ is kept, where
$m_{J/\psi}$ is the nominal $J/\psi$ mass~\cite{pdg2012}.  The fastest
positive and negative tracks are taken as the lepton candidates.  The
lepton species are identified with their $E/p$ ratios, where $E$ is
the measured energy deposition in the EMC of each track and $p$ is its
measured momentum.  The events with both $[E/p]^{+}<0.26$ and
$[E/p]^{-}<0.26$ are taken as $\mu^{+}\mu^{-}$ events, and those with
$[E/p]^{+}>0.80$, $[E/p]^{-}>0.80$, or $\sqrt{ ([E/p]^{+}-1)^2 +
([E/p]^{-}-1)^2}<0.4$ are taken as $e^{+}e^{-}$ events.  The backgrounds,
such as $J/\psi \to \pi^{+}\pi^{-}\pi^{0}$, are removed by requiring
the cosine of the angle between two lepton candidates be less than
$-0.95$.  The invariant mass of the lepton pair must be consistent
with that of a $J/\psi$, i.e., $M_{e^{+}e^{-}} \in [2.7,3.2]$
GeV$/c^{2}$ or $M_{\mu^{+}\mu^{-}} \in [3.0,3.2]$ GeV$/c^{2}$, where
different mass windows are used since the $e^{+}e^{-}$ final state has
more final state radiation than $\mu^{+}\mu^{-}$ does. Fig.~\ref{fig:ppjll}
shows the invariant masses of the dipion pair (top) and the dilepton
(bottom) pairs for $\pi^+\pi^-e^+e^-$ (left) and
$\pi^+\pi^-\mu^+\mu^-$ (right) final states.  To extract $N_{ll}$, we
count the number of events directly in a narrower mass window of
$M^{\rm rec.}_{\pi\pi}$.  Fig.~\ref{fig:ppjxx} shows the distributions of
the invariant mass recoiling against the dipion for the $e^{+}e^{-}$ ( top
right ) and $\mu^{+}\mu^{-}$ ( bottom right ) channels for the
$\pi^{+}\pi^{-}l^{+}l^{-}$ candidates.

\section{Background study}
For the $\pi^{+}\pi^{-}J/\psi$ final state, the backgrounds are
studied with the $\psi(3686)$ inclusive MC and the continuum data
sample.  The backgrounds can be classified into three categories: (1)
the non-$\pi^+\pi^-J/\psi$ decays of $\psi(3686)$, such as $\psi(3686)
\to $ light hadrons or $\psi(3686) \to \eta J/\psi$; (2) the
$\psi(3686) \to \pi^+\pi^-J/\psi$ decays, but one or both soft pions
are from $J/\psi$ decays; and (3) other backgrounds, including the
continuum process in $e^{+}e^{-}$ annihilation, beam-related, and
cosmic ray backgrounds.  As shown in the left plot of
Fig.~\ref{fig:ppjxx}, the backgrounds from the
non-$\pi^{+}\pi^{-}J/\psi$ and non-$\psi(3686)$ events are smooth and
produce no peak at the $J/\psi$ mass.  The second kind of background is
studied with toy MC simulation in which the contributions with one or
two charged tracks from $J/\psi$ decays are studied.  The background
shape is also found to be smooth with no peak at $J/\psi$ mass.

After all the requirements described above, the
$\pi^{+}\pi^{-}l^{+}l^{-}$ event samples are rather clean.  In the
window of the invariant mass recoiling against the dipion
$[m_{J/\psi}-15, m_{J/\psi}+15]$ MeV$/c^{2}$, for the
$\pi^{+}\pi^{-}e^{+}e^{-}$ final state, the background level is
estimated to be less than 0.10\%.  The largest background is
$\psi(3686)\rightarrow\eta J/\psi,
\eta\rightarrow\gamma\pi^{+}\pi^{-}, J/\psi\rightarrow e^{+}e^{-}$
($\sim0.04\%$).  and the second largest background is
$\psi(3686)\rightarrow \pi^{+} \pi^{-} J/\psi, J/\psi\rightarrow
\pi^{+}\pi^{-}\pi^{0}$ ($\sim0.03\%$).  For the
$\pi^{+}\pi^{-}\mu^{+}\mu^{-}$ final state, the total background level
is found to be 0.15\%.  The largest background is from $\psi(3686) \to
\pi^+\pi^-J/\psi$, $J/\psi \to \pi^+\pi^-$ ($\sim0.09\%$), and the
second largest background is $\psi(3686) \to \eta J/\psi$, $\eta \to
\gamma \pi^{+}\pi^{-}, J/\psi \to \mu^{+}\mu^{-}$ ($\sim0.02\%$).
Since the dominant backgrounds are exclusively simulated and
subtracted from the signal region according to the known branching
fractions and the scaled continuum data is subtracted, the remaining
background is only $0.03~(0.04) \%$ for the $e^{+}e^{-}
(\mu^{+}\mu^{-})$ channel.

\section{\bf Data analysis \label{analysis}}

\begin{figure}[htbp]
\centering
\includegraphics[width=9cm,height=9cm]{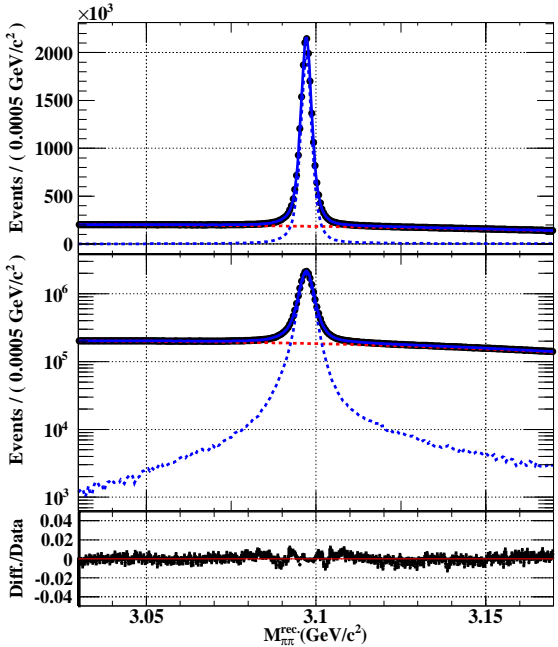}
\caption{ The dipion recoil mass spectrum for $\psi(3686) \to
\pi^{+}\pi^{-}J/\psi,~J/\psi \to \mbox{anything}$.  Top: data points
(black) overlaid with the fit result (solid blue curve) obtained using
the signal shape from $\psi(3686) \to \pi^{+}\pi^{-}J/\psi,~J/\psi \to
e^{+}e^{-}$ (blue dashed curve) and a second-order polynomial
background shape (red dashed curve).  Middle: the same plot as the top
but with a log scale.  Bottom: the fractional difference between the
fit and the data.}
\label{fig:fit2pipiJpsi}
\end{figure}

Since the dipion emission occurs independently of the subsequent
$J/\psi$ decay, the dipion recoil mass shape can be taken from any
cleanly determined $J/\psi$ decay. % This grants us considerable
%freedom from the accuracy of MC simulation in modeling the momentum
%resolution.
We use $J/\psi \to e^{+}e^{-}$, which is almost background-free and
has less background than $J/\psi \to \mu^{+}\mu^{-}$, for the signal
shape of the dipion recoil mass distribution, and use a second-order
polynomial to model the background shape.  Increasing the order of the
polynomial does not substantially improve the fit.  However, a study
shows that the resolution of the $\pi^{+}\pi^{-}$ recoil mass depends
on the charged track multiplicity of $J/\psi$ decays.  As a result,
the mass resolution from leptonic exclusive decays of $J/\psi$ is
slightly better than that of $J/\psi$ inclusive decays, and the
difference produces a bad fit quality ($\chi^{2}/ndof\sim 50$, where
$ndof$ is the number of degrees of freedom).  To improve the fit
quality , the signal shapes are smeared by convoluting them with two
Gaussian functions, whose parameters are determined by directly
fitting to data. While this procedure obviously improves the quality
($\chi^{2}/ndof \sim 4$), it changes the resultant ${\cal B}_{\pi\pi\psi}$ by
only 0.37\%, which is taken as one of the sources of systematic
uncertainty.  Fig.~\ref{fig:fit2pipiJpsi} shows the fit to the
dipion recoil mass spectrum for $\psi(3686) \to
\pi^{+}\pi^{-}J/\psi,~J/\psi \to \mbox{anything}$.

\begin{figure}[htbp]
\centering
\includegraphics[width=9cm,height=9cm]{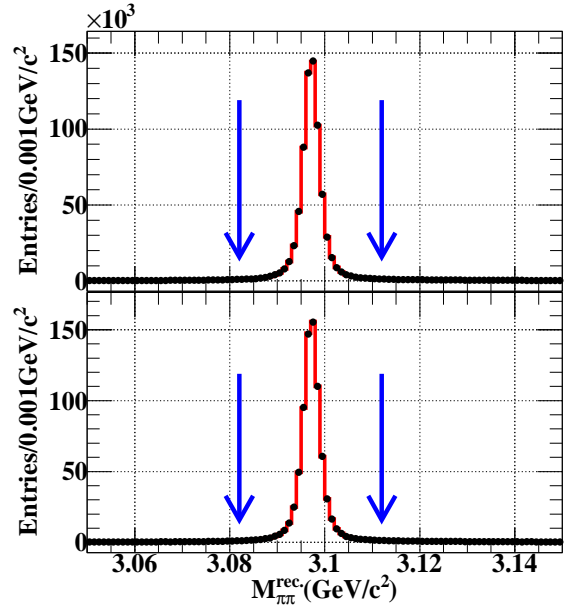}
\caption{ The dipion recoil mass spectrum for $\psi(3686) \to
\pi^{+}\pi^{-}J/\psi,~J/\psi \to l^{+}l^{-}$, the data points (black
dots) overlaid with the smeared MC simulation (solid red histogram)
according to the signal shape of data.  The regions between the
arrows are used to count the number of candidates.  top: $J/\psi
\to e^{+}e^{-}$; bottom: $J/\psi \to \mu^{+}\mu^{-}$.}
\label{fig:countjll}
\end{figure}

For the $\pi^{+}\pi^{-}l^{+}l^{-}$ final states, the number of signal
candidates in the distribution of $M^{\rm rec.}_{\pi^{+}\pi^{-}}$ are
counted directly, since they are almost background free.  However, as
shown in the right column of Fig.~\ref{fig:ppjxx}, the resolutions of
data (black dots) and MC simulation (red histogram) are
different. Thus, the MC distributions are smeared according to data in
determining their reconstruction efficiencies.  A mass window of
$[m_{J/\psi}-15, m_{J/\psi}+15]$ MeV$/c^{2}$ ($\sim5\sigma$) is used
in counting the signal candidates.  Fig.~\ref{fig:countjll} shows the
comparison between data and the smeared MC simulation, in which the
data points, as well as the regions, are the same as those in the
right panel of Fig.~\ref{fig:ppjxx}.

%\subsection{Input/output checks}
To validate the analysis method, MC input/output checks are performed
based on the $106 \times 10^{6}$ $\psi(3686)$ inclusive MC sample,
which has input values ${\cal B}_{\pi\pi\psi}$, ${\cal B}_{ee}$, and
${\cal B}_{\mu\mu}$ of 32.6\%, 5.93\%, and 5.94\%, respectively.
Since this sample can not be used at the same time to determine the
efficiencies, an alternative $10^{7}$ $\psi(3686)$ inclusive MC sample
is used for their determination. In order to make these two samples
look more like real data, we also add in the scaled continuum data.
As shown in Table~\ref{tab:summary-io-results}, all the extracted
branching fractions are consistent with the input branching fractions
within their uncertainties.

\begin{table}[htbp] 
\caption{Summary of MC input/output check results of the three
  processes (${\cal B }$ is in percent).
}
\label{tab:summary-io-results}
\begin{center}
\begin{small}
\begin{tabular}{lccc}\hline\hline
     modes                      &${\cal B}_{\rm in}$&$N_{\rm obs}(10^{3})$  &${\cal B}$\\\hline
$\pi^{+}\pi^{-}J/\psi$          &32.6 &$18783.4\pm  5.1$ &$32.64\pm 0.03$ \\
$\pi^{+}\pi^{-}e^{+}e^{-}$      & 5.93 &$660.6   \pm  0.8   $  &$5.912\pm 0.024$ \\
$\pi^{+}\pi^{-}\mu^{+}\mu^{-}$  & 5.94 &$707.5   \pm  0.8 $    &$5.930\pm 0.024$ \\
\hline\hline
\end{tabular}
\end{small}
\end{center}
\end{table}

Table~\ref{tab:summary-of-results} summarizes the resultant signal
yields, efficiencies, and branching fractions based on data, along
with their statistical uncertainties.

\begin{table}[htbp] \begin{center}
%\begin{table}\begin{widetext}
\caption{ Summary of $\psi(3686) \to \pi^{+}\pi^{-}J/\psi$ and
$J/\psi\to l^{+}l^{-}$ results, showing numbers of the three decays,
$N_{\pi\pi\psi}$, $N_{ee}$ and $N_{\mu\mu}$; efficiencies for
observing those decays, $\epsilon_{\pi\pi\psi}$, $\epsilon_{ee}$ and
$\epsilon_{\mu\mu}$; and the calculated branching fractions of the
three channels, along with the statistical uncertainties on all
quantities.  }
\label{tab:summary-of-results}
\begin{small}
%\begin{tabular}{l|ccc}\hline\hline    
%                               & $N(10^{3})$        & $\epsilon$(\%)      & ${\cal B}$($\%$)  \\\hline
%$\pi^{+}\pi^{-}J/\psi$         & $20235 \pm 6 $     & $54.37\pm 0.02$     & $34.98\pm  0.02$ \\ 
%$\pi^{+}\pi^{-}e^{+}e^{-}$     & $718.8 \pm 0.9$    & $32.19\pm 0.04$     & $5.983\pm 0.007$ \\       
%$\pi^{+}\pi^{-}\mu^{+}\mu^{-}$ & $771.1 \pm 0.9$    & $34.54\pm 0.04$     & $5.973\pm 0.007$\\  
%\hline\hline
%\end{tabular}
\begin{tabular}{l|ccc}\hline\hline    
                 & $\pi^{+}\pi^{-}J/\psi$     & $\pi^{+}\pi^{-}e^{+}e^{-}$& $\pi^{+}\pi^{-}\mu^{+}\mu^{-}$\\\hline
$N(10^{3})$      &  $20235 \pm 6 $            & $718.8     \pm 0.9$       &    $771.1     \pm 0.9$        \\
$\epsilon$(\%)   &  $54.37\pm 0.02$           & $32.19     \pm 0.04$      &    $34.54     \pm 0.04$       \\
${\cal B}$($\%$) &  $34.98\pm  0.02$          & $5.983     \pm 0.007$     &    $5.973     \pm 0.007$      \\
\hline\hline
\end{tabular}
\end{small}
\end{center}
\end{table}

\section{\bf Study of systematic uncertainties\label{systematic_error}}
We consider systematic uncertainties from many different sources.  The
uncertainty of the number of $\psi(3686)$ decays,
0.81\%~\cite{ref:psiptotnumber}, which is measured by counting the
hadronic events from $\psi(3686)$ decay directly, is the dominant
uncertainty of ${\cal B}_{\pi\pi\psi}$, while ${\cal B}_{ee}$ and
${\cal B}_{\mu\mu}$ are independent of it.  The difference of tracking
efficiency between data and MC simulation is measured from a
comparison of yields of partially and fully reconstructed $\psi(3686)
\to \pi^{+}\pi^{-}J/\psi$ and $J/\psi \to l^{+}l^{-}$ decays in real
and simulated data.  The differences depending on the polar angle and
the transverse momentum of the track are used to re-weight the MC
samples.  And the uncertainty of the re-weighting factor is estimated
to be 0.1\% per lepton and 0.4\% per pion. The systematic effects
related to the soft pion tracking cancel in the calculation of ${\cal
B}_{ee}$ and ${\cal B}_{\mu\mu}$.  The tracking uncertainties of
$\pi^{+}$ and $\pi^{-}$, or $l^{+}$ and $l^{-}$ are considered as
fully correlated and are added linearly.

In the inclusive analysis, even though we only reconstruct two soft
charged pions, the reconstruction efficiency depends on the track
multiplicity of the subsequent $J/\psi$ decays.  However, since the
sum of known exclusive $J/\psi$ partial widths is small compared to
the total width, a MC sample must be used to represent all $J/\psi$
decays and to obtain $\epsilon_{\pi\pi\psi}$.  The global efficiency
found is about $54\%$ but varies about 15\% (relative) from low to high
charged track multiplicities of $J/\psi$ decays, similar to that
reported in BESI~\cite{bes:jll}, but the variation is much larger than
that in CLEO-c~\cite{cleo:ppjll}.  We attribute the difference to the
finer segmentation in the CLEO-c tracking system, which was designed
for physics at higher energy \cite{ref:cleo-detector} relative to that
of BESI~\cite{ref:bes1-detector} and BESIII~\cite{ref:bes3}, as well
as the consequent robustness of track reconstruction in the presence
of many charged particles.

\begin{figure}[htbp]
\centering
\includegraphics[width=8cm,height=6cm]{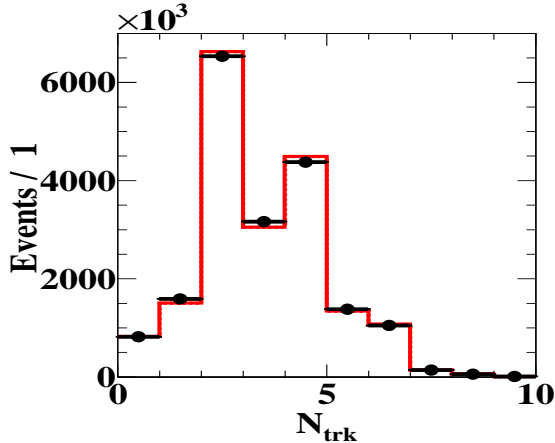}
\caption{
%Fit to the multiplicity distribution of data with that of the MC sample
Fit (histogram) to the multiplicity distribution of data (points) with that of the MC sample
}
\label{fig:fitmultiplicity}
\end{figure}

To study the dependence of the detection efficiency
$\epsilon_{\pi\pi\psi}$ on the generated charged track multiplicity
distribution for $J/\psi$ decays in $\pi^{+}\pi^{-}J/\psi$ events, we
first use the inclusive MC sample to determine the detection
efficiency ($\epsilon_k$) as a function of generated track
multiplicity ($k$), as shown in Table~\ref{tab:summary-of-106M}, and
then determine $\epsilon_{\pi\pi\psi}$ considering alternative
generated multiplicity distributions. Two methods are used to
determine the fraction $w_{k}$ of each multiplicity from data directly
and $\epsilon_{\pi\pi\psi}$. The first is the method used in
Ref.~\cite{bes:jll}, which fits the observed multiplicities in data
using the efficiency matrix, $\epsilon_{ij}$, which describes the
efficiency of a MC event generated with $j$ charged tracks to be
reconstructed with $i$ charged tracks, to determine
the true generated charged track multiplicity distribution.  The
second method fits the observed multiplicity distribution with
exclusive MC based templates as in
Ref.~\cite{cleo:ppjll}. Fig.~\ref{fig:fitmultiplicity} shows the
multiplicity distribution fitted by the generated multiplicity
distribution of the inclusive MC.  Table~\ref{tab:summary-of-106M}
summarizes the multiplicity distribution obtained from the
$\psi(3686)$ inclusive MC and the two methods mentioned above, as well
as the overall $\epsilon_{\pi\pi\psi}$ for each case.  Consistent
results are obtained, which indicates that $\epsilon_{\pi\pi\psi}$ is
not very sensitive to the generated multiplicity distribution of
$J/\psi$ decays. We assign the largest difference as the systematic
uncertainty due to our imperfect simulation of the charged track multiplicity, 
$N_{\rm trk}^{J/\psi}$, in $J/\psi$ decays.

%To study the dependence of the detection efficiency on the charged
%track multiplicity for $J/\psi$ decays in $\pi^{+}\pi^{-}J/\psi$ events, we
%choose an appropriate mixture of $J/\psi$ decay modes from MC
%simulation to fit the data, and determine the fraction ($w_{k}$) of
%each multiplicity from data directly.  Two methods are used to
%determine $w_{k}$ and $\epsilon_{\pi\pi\psi}$.  One is the method used
%in Ref.~\cite{bes:jll}, which is an unfolding method, working on the
%observed multiplicities in data and the efficiency matrix from MC
%simulation.  The other is similar to the method in
%Ref.~\cite{cleo:ppjll}, but the MC simulated multiplicity
%distributions tagged from $\psi(3686)$ inclusive MC are used instead of
%those in Ref.~\cite{cleo:ppjll}, which are from some special MC
%samples.  Fig.~\ref{fig:fitmultiplicity} shows the multiplicity
%distribution fitted by the multiplicity distribution of the MC.  %%?????
%Table~\ref{tab:summary-of-106M} summarizes the results obtained from the
%$\psi(3686)$ inclusive MC and the two methods mentioned above.
%Consistent results are obtained, which indicates that
%$\epsilon_{\pi\pi\psi}$ is not sensitive to the multiplicity
%distribution of $J/\psi$ decays.

\begin{table}[htbp] 
\caption{ The fractions of each charged track multiplicity of $J/\psi$
decays from the $\psi(3686)$ inclusive MC (column 2), from the method
of Ref.~\cite{bes:jll} (column 3), and that of Ref.~\cite{cleo:ppjll}
(column 4).  The MC efficiency for $k$-charged tracks is shown as
$\epsilon_{k}$. The overall efficiency $\epsilon_{\pi\pi\psi}$ for
each of the three cases is also shown.}
\label{tab:summary-of-106M}
\begin{center}
\begin{small}
\begin{tabular}{lcccc}\hline\hline
 $N^{J/\psi}_{\rm trk}$ & $w_k$            & $w_k$ & $w_k$
  &  $\epsilon_k (\%)$ \\
    &  (incl.)            & (BES) &  (CLEO-c) &  \\\hline
       0                       & 0.0175                       & 0.0225           & 0.0231              & 56.56  \\
       2                       & 0.3440                       & 0.3881           & 0.3945              & 55.82  \\
       4                       & 0.4310                       & 0.4015           & 0.4012              & 53.97  \\
       6                       & 0.1871                       & 0.1644           & 0.1627              & 52.03  \\
       8                       & 0.0199                       & 0.0200           & 0.0185              & 49.49  \\
$\epsilon_{\pi\pi\psi}$ (\%) &  54.17   &  54.15    &  54.36    &   \\
\hline\hline
\end{tabular}
\end{small}
\end{center}
\end{table}

From the above analysis, the uncertainty from the charged track
multiplicity distribution was found to be less than 0.2\%, including
all the contributions from the fit, the sideband selection, and the
backgrounds.  The efficiency does exhibit a weak dependence not only
on the charged multiplicity, but also slightly on the neutral track
multiplicity.  More neutral particles in the $J/\psi$ decay soften the
momentum spectrum of the charged tracks, which makes the tracks harder
to detect, and produces more photon conversions in the material in the
inner detectors, which also changes the charged track multiplicity.
But a MC study suggests that such effects are very small and can be
neglected.

The dipion invariant mass distribution is simulated with the
measurement of Ref.~\cite{ref:bes1ppjdistribution}, in which a small
amount of $D$-wave contribution is included.  However, there is still
a slight difference between the data and MC simulation, so the MC
simulation is re-weighted by the distribution in data, and the
difference before and after the re-weighting, which is 0.35\% for
$\epsilon_{\pi\pi\psi}$, is taken as a systematic uncertainty.  The
difference is much smaller for $\epsilon_{l^{+}l^{-}}$, $\sim0.01\%$,
since the effect cancels in a relative measurement.

The fit to the huge statistics of the distribution of mass recoiling
against the dipion gave a poor $\chi^{2}/ndof$, since the resolutions
in the exclusive and inclusive decays are a bit different.  The signal
shapes of the exclusive channel are smeared by convoluting with double
Gaussian functions to improve the fit quality.  And as a result,
${\cal B}_{\pi\pi\psi}$ is changed by 0.37\% before and after the
smearing, which is taken as one of the systematic uncertainties.

The shapes of the invariant mass distributions of lepton pairs are
affected by the simulation of final state radiation (FSR), which is
simulated with the PHOTOS package~\cite{ref:photos}.  Differences
between data and MC simulation are still observed.  The invariant mass
requirement on lepton pairs is studied by an alternative control
sample, in which the lepton pairs are identified by the information of
the EMC, MUC, and specific ionization ($dE/dx$) measured in MDC, while
demanding $M^{\rm rec.}_{\pi^{+}\pi^{-}}$ to be consistent with the
$J/\psi$ mass, but without any requirement on the invariant mass of
$e^{+}e^{-}$ or $\mu^{+}\mu^{-}$.  The differences are determined to
be 0.29\% ($e^{+}e^{-}$) and 0.45\% ($\mu^{+}\mu^{-}$).  To reduce
this type of uncertainty, corrections are made based on this study,
and the final contributions to the total systematic uncertainty are
0.10\% and 0.23\%, respectively.

The remaining sources of systematic uncertainty not addressed above
are the requirements on $E/p$, the angles between the two leptons and
the two pions, the background contamination for
$\pi^{+}\pi^{-}l^{+}l^{-}$ final states, and the uncertainty related
to the fitting (counting) procedure.  The first two items are
determined with independent samples selected with alternative
selection criteria, and the uncertainties of the $E/p$ requirement are
found to be 0.18\% and 0.09\% for muon and electron pairs,
respectively; the uncertainties of the two angle requirements are
found to be less than 0.1\%.  The uncertainties of the backgrounds of
the $\pi^{+}\pi^{-}l^{+}l^{-}$ exclusive final states are only
0.03$\sim$0.04\%, after subtracting the background using known
branching ratios.  The uncertainties of the fitting (except the
uncertainty of the resolution in ${\cal B}_{\pi\pi\psi}$), which are
all at the part per thousand level, are estimated by changing the
signal shape, background shape, fitting ranges (mass windows), and bin
size.  The uncertainties of the trigger efficiency in the three
measurements are taken as 0.10\% for ${\cal B}_{\pi\pi\psi}$ and
0.30\% for ${\cal B}_{ll}$ according to the study in
\cite{ref:trigger}.

The systematic uncertainties in the branching fractions are summarized
in Table~\ref{table::system-of-3channels}.  The systematic uncertainty
in ${\cal B}[\psi(3686) \to \pi^+\pi^- J/\psi]$ is dominated by the
number of $\psi(3686)$ events and the tracking efficiency of the
two soft pions, and the total contribution of the other sources is
less than 0.5\%.  The systematic uncertainty in ${\cal B}[J/\psi\to
l^+l^-]$ is dominated by the uncertainty of the determination of
$N_{\pi\pi\psi}$.
%by the background contamination, which are at per mille level. 
%More strict selection criteria can be used to suppress the background level to less than 0.1\%, 
%but this will cause large systematic uncertainty due to imperfect MC simulation.

%%%
\begin{table}[htbp] 
\caption{Summary of the systematic uncertainties (\%) in the branching fractions. }
\label{table::system-of-3channels}
\begin{center}
\begin{small}
\begin{tabular}{l|ccc}\hline\hline
Sources & $\pi^{+}\pi^{-}J/\psi$ & $e^+ e^-$ &$\mu^+\mu^-$ \\\hline
Tracking                            &  0.80  &  0.20   &  0.20\\
Multiplicity of $J/\psi$            &  0.20  &  0.20   &  0.20 \\
$M_{\pi^+\pi^-}$ distribution       &  0.35  &  0.01   &  0.01\\
Background shape                    &  0.03  &  0.03   &  0.04\\
Fit/Count range                     &  0.06  &  0.14   &  0.14\\
Bin size                            &  0.06  &  0.06   &  0.06\\
$E/p$                               &   ---  &  0.18   &  0.09 \\
$\cos\theta_{\pi^+\pi^-}$           &  0.13  &  0.07   &  0.07\\
$\cos\theta_{l^+l^-}$               &   ---  &  0.04   &  0.05\\
FSR effect of $l^+l^-$              &   ---  &  0.10   &  0.23\\
Fit method                          &  0.37  &  0.37   &  0.37\\
Trigger                             &  0.10  &  0.30   &  0.30\\
Number of $\psi(3686)$              &  0.81  &  ---    &  --- \\ \hline
Sum in quadrature                   &  1.28  &  0.62   &  0.63\\
\hline\hline
\end{tabular}
\end{small}
\end{center}
\end{table}

\section{\bf  summary  and discussion \label{summary}}
%\bigskip

The branching fractions of three processes $\psi(3686) \to \pi^{+} \pi^{-}
J/\psi$, $J/\psi \to e^{+} e^{-}$, and $J/\psi \to \mu^{+} \mu^{-}$,
are measured with $(106.41\pm 0.86)\times 10^6$ $\psi(3686)$ decays.
The results are ${\cal B}_{\pi\pi\psi} =(34.98\pm 0.02 \pm 0.45)\%$,
${\cal B}_{ee} = (5.983\pm 0.007 \pm 0.037 )\% $, and ${\cal
B}_{\mu\mu} = (5.973\pm 0.007 \pm 0.038 )\%$, where the first
uncertainties are statistical and the second are systematic.  We also
measure ${\cal B}_{ee}/{\cal B}_{\mu\mu} = 1.0017 \pm 0.0017 \pm
0.0033 $, where the common systematic uncertainties have been canceled
out.  This tests $e$-$\mu$ universality at the four tenths of a
percent level.  The precision is significantly improved with
respective to the PDG average ${\cal B}_{ee}/{\cal B}_{\mu\mu} = 0.998
\pm 0.012$~\cite{pdg2012}.  Assuming leptonic universality, the
average of ${\cal B}_{ee}$ and ${\cal B}_{\mu\mu}$ is ${\cal
B}[J/\psi\to l^+l^-]=(5.978\pm 0.005\pm 0.040)\%$, in which the
correlations among the uncertainties are accounted for.  The measured
branching fractions of $J/\psi\to e^+e^-/\mu^+\mu^-$ are consistent
with previous measurements, and will allow improvements in potential
models~\cite{potentials} and the determinations of $\Gamma_{ee}$ and
$\Gamma_{\rm tot}$ of $J/\psi$~\cite{babar}.

\begin{figure}[htbp]
\centering
\includegraphics[width=7cm,height=6cm]{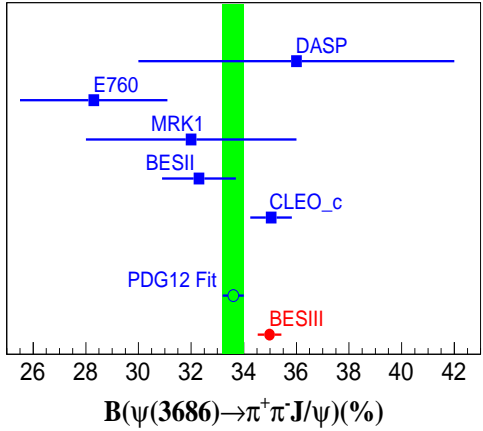}
\caption{
Comparison of ${\cal B}[\psi(3686)\to \pi^{+}\pi^{-}J/\psi]$ among different experiments.}
\label{fig:compare-Bppj}
\end{figure}

Figure~\ref{fig:compare-Bppj} shows 
a comparison of  ${\cal B}_{\pi\pi\psi}$ among various experiments. 
Our measured ${\cal B}_{\pi\pi\psi}$ is the most precise to date and is consistent with the latest CLEO-c~\cite{cleo-ppj} measurement, 
but higher than most of the previous measurements.
 
%For the branching fractions of  $J/\psi \to e^+ e^-$  and $J/\psi \to \mu^+\mu^-$,
%the central values are consistent with previous measurements, 
%\bigskip

\section{\bf ACKNOWLEDGMENTS}

The BESIII collaboration thanks the staff of BEPCII and the computing center for their strong support. 
This work is supported in part by the Ministry of Science and Technology of China under Contract No. 2009CB825200; 
National Natural Science Foundation of China (NSFC) under Contracts Nos. 
10625524, 10821063, 10825524, 10835001, 10935007, 11125525, 11235011, 11005115; 
Joint Funds of the National Natural Science Foundation of China under Contracts Nos. 11079008, 11179007, 10979058; 
the Chinese Academy of Sciences (CAS) Large-Scale Scientific Facility Program; 
CAS under Contracts Nos. KJCX2-YW-N29, KJCX2-YW-N45; 
100 Talents Program of CAS; 
German Research Foundation DFG under Contract No. Collaborative Research Center CRC-1044; 
Istituto Nazionale di Fisica Nucleare, Italy; 
Ministry of Development of Turkey under Contract No. DPT2006K-120470; 
U.S. Department of Energy under Contracts Nos. DE-FG02-04ER41291, DE-FG02-05ER41374, DE-FG02-94ER40823; 
U.S. National Science Foundation; 
University of Groningen (RuG) and the Helmholtzzentrum fuer Schwerionenforschung GmbH (GSI), Darmstadt; 
WCU Program of National Research Foundation of Korea under Contract No. R32-2008-000-10155-0.

\end{document}